\documentclass[%
superscriptaddress,
amsmath,amssymb,
aps,
floatfix,
]{revtex4-2}

\usepackage{graphicx}
\usepackage{bm}

\usepackage{amsfonts}
\begin{document}


\title{Photonic Elementary Cellular Automata for Simulation of Complex Phenomena}

\author{Gordon H.Y. Li}
\affiliation{Department of Applied Physics, California Institute of Technology, Pasadena, CA 91125, USA}
\author{Christian R. Leefmans}
\affiliation{Department of Applied Physics, California Institute of Technology, Pasadena, CA 91125, USA}
\author{James Williams}
\affiliation{Department of Electrical Engineering, California Institute of Technology, Pasadena, CA 91125, USA}
\author{Alireza Marandi}
\email{marandi@caltech.edu}
\affiliation{Department of Applied Physics, California Institute of Technology, Pasadena, CA 91125, USA}
\affiliation{Department of Electrical Engineering, California Institute of Technology, Pasadena, CA 91125, USA}


\begin{abstract}
Cellular automata are a class of computational models based on simple rules and algorithms that can simulate a wide range of complex phenomena. However, when using conventional computers, these `simple' rules are only encapsulated at the level of software. This can be taken one step further by simplifying the underlying physical hardware. Here, we propose and implement a simple photonic hardware platform for simulating complex phenomena based on cellular automata. Using this special-purpose computer, we experimentally demonstrate complex phenomena including fractals, chaos, and solitons, which are typically associated with much more complex physical systems. The flexibility and programmability of our photonic computer presents new opportunities to simulate and harness complexity for efficient, robust, and decentralized information processing schemes using light.
\end{abstract}

\maketitle
Modern digital electronic computers based on the von Neumann architecture, composed of billions of transistors engineered in a hierarchical and highly structured manner, engender extreme hardware complexity in their construction. Unlike the von Neumann architecture, nature is abound with emergent phenomena and complex systems containing many interacting components following simple rules with no hierarchical control. For example, social insects like ants with only limited local information can collectively self-organize to form global structures~\cite{bonabeau1997self}. This suggests that an alternative, and potentially more efficient, way to simulate such phenomena is to harness simple and decentralized physical hardware that directly emulates the underlying rules of a complex system.

One class of computational models that can benefit from simple and decentralized physical hardware are Cellular Automata (CA), which exhibit complex behaviour emerging from the local interactions of cells arranged on a regular lattice~\cite{wolfram1984cellular}. CA were introduced in the 1940s to study how self-replication and evolution can emerge in artificial life~\cite{neumann1966theory}, and as later popularized in Conway's Game of Life~\cite{gardner1970fantastic}, which exhibits self-organizing patterns reminiscent of biological systems. Subsequent landmark studies revealed that CA are also capable of replicating other complex behaviour such as fractals~\cite{wolfram1983statistical}, chaos~\cite{wolfram1986random}, self-organized criticality~\cite{bak1987self}, synchronization~\cite{balzer19678}, and universal computation~\cite{cook2004universality}. Consequently, CA have found utility in modelling a wide range of natural phenomena in physics~\cite{wolf2004lattice,chopard1991cellular}, chemistry~\cite{gerhardt1989cellular,ashwin2008fast,raabe2002cellular}, and biology~\cite{ermentrout1993cellular}. Furthermore, CA have important applications in real-world computational problems such as cryptography~\cite{wolfram1985cryptography}, data compression~\cite{lafe1997data}, error-correction~\cite{chowdhury1994design}, and developing more robust artificial intelligence~\cite{mordvintsev2020growing}. Owing to their simple formulations, certain CA of interest are computationally irreducible~\cite{israeli2004computational}, i.e. there are no analytical shortcuts to perform the effective calculation without resorting to executing the sequential simulation in its entirety. On the other hand, most CA are only implemented as high-level software on conventional computers, resulting in unnecessary overhead.  Therefore, it is desirable to seek out physical hardware that better encapsulates the computational principles of CA to enable more efficient simulation. Notable previous attempts to implement physical systems tailored to perform CA include self-assembling DNA molecules~\cite{rothemund2004algorithmic}, arrays of nanomagnets~\cite{imre2006majority}, memristor networks~\cite{itoh2019memristor}, and living slime moulds~\cite{shirakawa2015construction}.

In this work, we propose and experimentally implement a photonic computational platform capable of simulating complex phenomena using CA. Compared to other approaches, our photonic platform offers several distinct advantages: (1) the inherently high-bandwidth and parallelism endowed by computing using light offers potentially orders-of-magnitude speed-up in clock rate over the simulation of CA on conventional von Neumann computers, (2) rapid reconfigurability for easy programming of a variety of CA rules enables many different complex phenomena to be observed in the \textit{same} physical system, and (3) the kind of sparse, local, and shift-invariant connections required for CA are well-suited for this platform. We will demonstrate how even simple photonic hardware can host a wide range of complex emergent phenomena and is capable of sophisticated (or even universal) computation. By exploiting this complexity, we reveal a path towards the next generation of more efficient or robust photonic hardware accelerators for reservoir computing~\cite{yilmaz2014reservoir,nichele2017deep} and deep learning~\cite{mordvintsev2020growing,randazzo2020self}.

\section*{Results}

\begin{figure}
\includegraphics{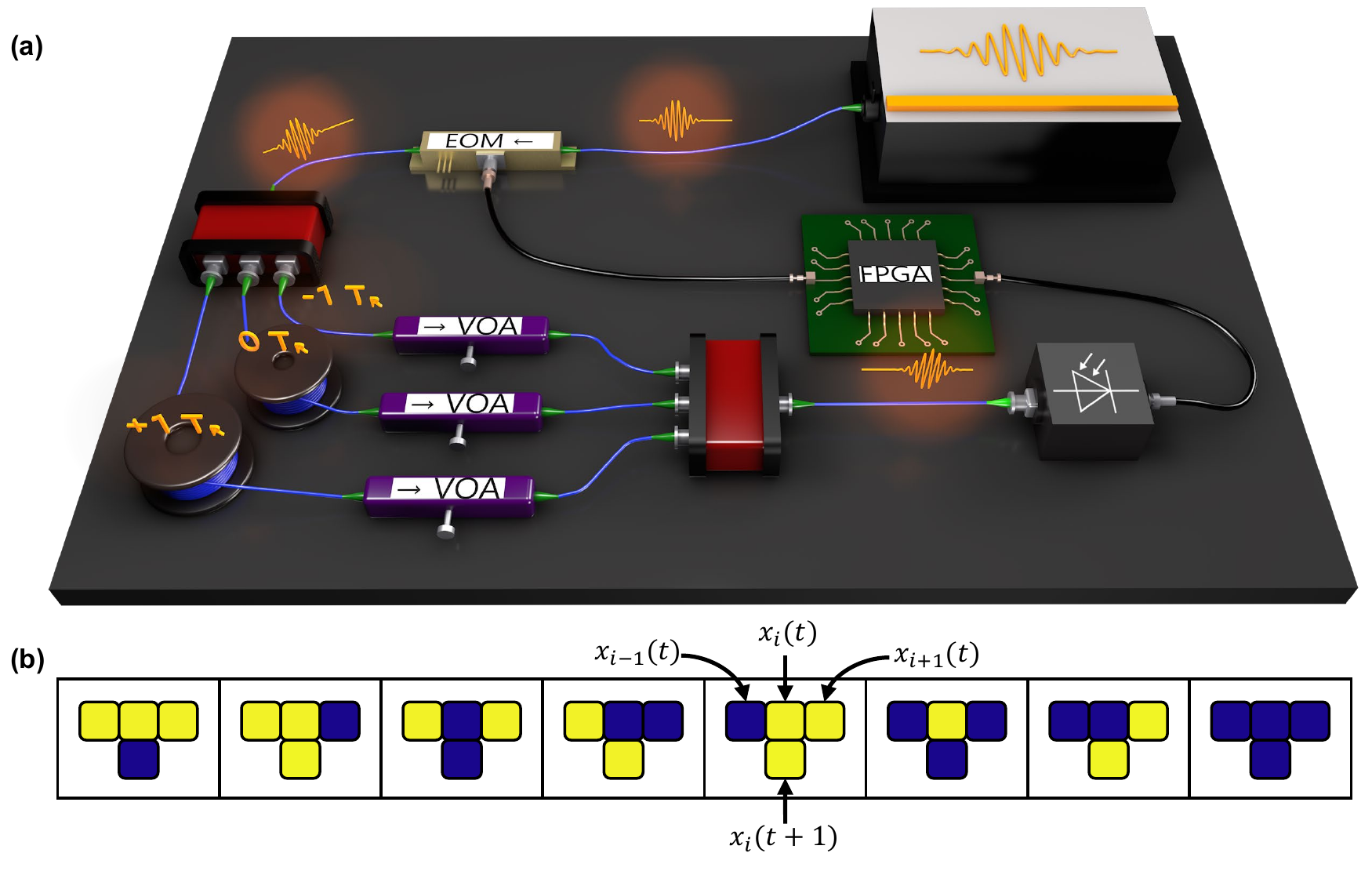}
\caption{\textbf{Photonic platform for simulating complex phenomena using Elementary Cellular Automata.} (a) Schematic of the experimental setup. Cells are represented by pulses of light produced by a laser with a repetition rate of $T_{R}$. The cell states are encoded by an electro-optic modulator (EOM) and are split into optical fiber delay lines (blue lines) to induce local interactions of neighbouring light pulses. Specific ECA rules are programmed by tuning the variable optical attenuator (VOA) in each delay line. Optoelectronic thresholding is performed following the coherent interference of light pulses, with the resultant cell states stored on a field-programmable gate array (FPGA) and reinjected (black lines) to drive the input EOM for the next iteration. (b) Truth table showing the uniform and synchronous update for ECA Rule 90 with the top row in each case representing the current states of the 3-cell neighbourhood and the bottom row showing the cell state during the next iteration.}
\label{fig:1}
\end{figure}
We focus on the simplest types of CA called Elementary Cellular Automata (ECA)~\cite{wolfram1983statistical}. These are discrete-time dynamical systems defined on a 1D lattice of cells with binary states that evolve according to Eq.~\ref{eq:1}:
\begin{equation}
    x_{i}(t+1)=f\left(x_{i-1}(t),x_{i}(t),x_{i+1}(t)\right)\ ,
    \label{eq:1}
\end{equation}
where $x_{i}(t)\in\mathbb{Z}_{2}$ is the state (i.e. dead or live) of the $i^{\mathrm{th}}$ cell at time step $t$, and $f:(\mathbb{Z}_{2})^{3}\rightarrow\mathbb{Z}_{2}$ is the update rule. Crucially, the rules specifying interactions amongst cells are computed using only local nearest-neighbour information without reference to the global pattern. Remarkably, not only can the underlying rules be simple, but the initial conditions can also be simple - consisting, for example, of just a single live cell - and yet the collective behaviour produced can still be highly complex~\cite{wolfram1983statistical}. The 256 possible ECA rules encapsulate a wide range of complex phenomena and are representative of the four universality classes of increasing complexity introduced by Wolfram~\cite{wolfram1984universality}.

Here, we implement ECA in a time-multiplexed photonic system as shown in Fig.~\ref{fig:1}(a). Cell states are represented using pulses of light produced by a laser with a fixed repetition rate $T_{R}$; the presence of a pulse indicates a live cell, and dead otherwise. In this way, the 1D lattice sites correspond to
time bins in a synthetic temporal dimension~\cite{leefmans2022topological,marandi2014network}, hence permitting a lattice that extends indefinitely with an arbitrary number of cells by time-multiplexing a single nonlinear node. The pulse amplitude/phase representing the initial cell state is encoded using an electro-optic modulator (EOM), then equally split into three paths. Nearest-neighbour cell interactions are achieved using coherent interference of the optical pulses passed through the delay lines with time delay $\pm T_{R}$ relative to the cell being updated (the optical pulse through the $0$ delay line), followed by optoelectronic thresholding to enforce the binary states. Finally, the optoelectronic signal is stored on a field-programmable gate array (FPGA), which performs feedback of the measured cell states by driving the input EOM for the next iteration. By repeating this process for many cycles, we observe the emergence of complex phenomena in the cell states of the ECA. The desired ECA update rule, such as ECA Rule 90 (following the Wolfram naming convention) shown in Fig.~\ref{fig:1}(b), is programmed by tuning the thresholding value and variable optical attentuator (VOA) in each delay line, which represents constant amplitude/phase weights. This rule encoding can be interpreted as a weighted linear summation followed by a nonlinear thresholding function, which is akin to a single perceptron in the context of artificial neural networks~\cite{lecun2015deep}. Therefore, the dynamics of the abstract ECA rule is exactly mapped to the physical time-evolution of the photonic simulator. 

\begin{figure}[t]
\includegraphics{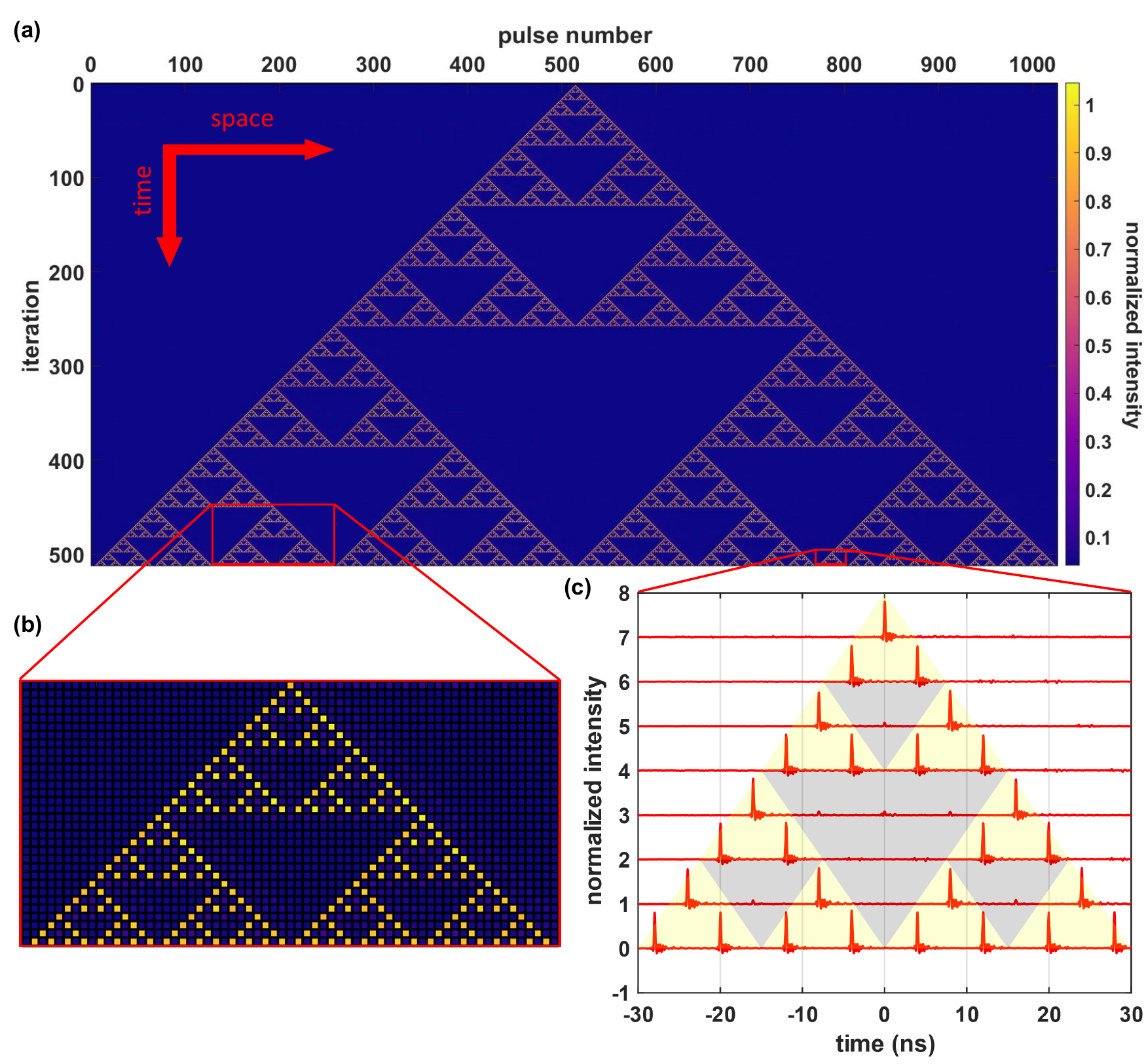}
\caption{\textbf{Experimental result of ECA Rule 90 on the photonic hardware starting from a single live cell.} (a) Zoomed-out equivalent space-time diagram showing emergence of the Sierpinski Triangle fractal. (b) Zoomed-in view showing the fractal self-similarity down to the cellular scale. (c) Time traces (vertically separated for easier viewing) of the individual light pulses representing each cell separated by $4~\mathrm{ns}$.}
\label{fig:2}
\end{figure}

Firstly, one of the most striking patterns that emerge in CA are fractals, which are often self-similar geometric shapes that appear the same at any scale. Fractals are ubiquitous in nature and occur in a diverse range of physical phenomena including the rings of Saturn~\cite{li2015edges}, snowflakes~\cite{libbrecht2005physics}, and fault geology~\cite{okubo1987fractal}. ECA Rule 90, defined in Fig.~\ref{fig:1}(b), provides a simple model for fractal formation and self-replication. The local update rule can be expressed succinctly in terms of Boolean algebra as $x_{i}(t+1)=x_{i-1}(t)\bigoplus x_{i+1}(t)$, where $\bigoplus$ denotes the exclusive-or (XOR) logical operation. Thus, for this specific rule, the iterated cell state depends only on the states of its two neighbours. The fractal pattern is an emergent property of the nonlinear dynamics in the photonic computer, rather than being imposed on the system by an external ordering influence such as explicit geometric constructions in previous studies of photonic fractals~\cite{xu2021quantum,yang2020photonic,biesenthal2022fractal}. We show the experimentally obtained space-time equivalent diagram of ECA Rule 90, starting from a single live cell, in Fig.~\ref{fig:2}(a). 
The position of a cell in space (left-to-right) is represented by the pulse number in the synthetic temporal dimension, and the discrete-time step (top-to-bottom) is defined according to Eq.~\ref{eq:1}. The color of each cell is determined by the normalized peak pulse intensity \textit{before} thresholding. We see that the space-time diagram is in the shape of the well-known Sierpinski Triangle. This fractal can be constructed by recursively subdividing an equilateral triangle into four smaller equilateral triangles and removing the central triangle. It is characterized by a non-integer Hausdorff or fractal dimension of $\log3/\log2\approx1.585$. The self-similarity of the fractal shape persists down to the cellular scale as shown in Fig.~\ref{fig:2}(b) and can be seen in the time traces, shown in Fig.~\ref{fig:2}(c), for individual light pulses representing each cell. In this case, the middle optical delay line ($0\ T_{R}$ in Fig.~\ref{fig:1}(a)) can be ignored since the iterated cell state does not depend on its current state. This allows for an excellent extinction ratio between pulse peaks for live and dead cells and indicates that the ECA Rule 90 is implemented as intended. 

\begin{figure}[b]
\includegraphics{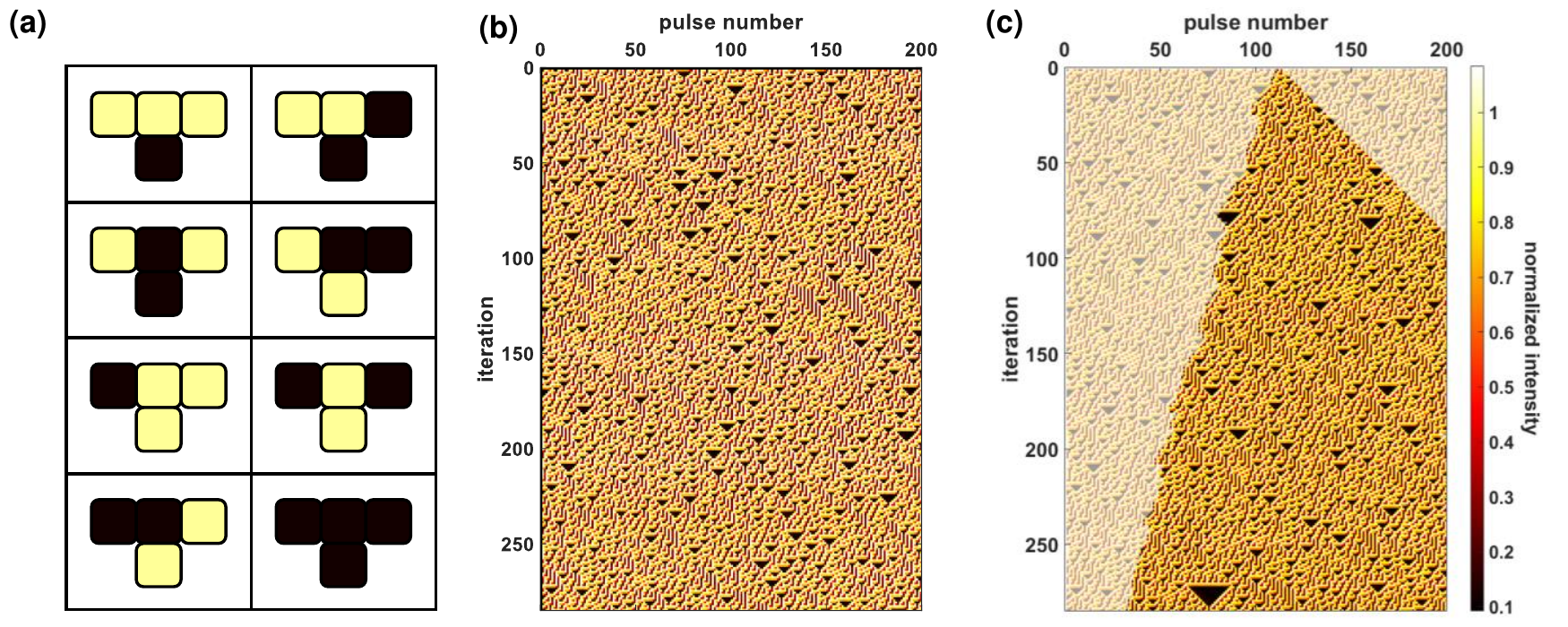}
\caption{\textbf{Chaotic patterns produced by ECA Rule 30 on the time-multiplexed photonic hardware.} (a) Truth table showing the update for ECA Rule 30. (b) Space-time diagram of ECA Rule 30 starting from a random initial condition showing chaotic dynamics. (c) Inverting a single cell state in the initial condition produces a pattern with differences that grow linearly to the right and asymptotically linearly to the left (regions that are different to (b) are highlighted, and regions that are identical are displayed as partially transparent), hence demonstrating sensitivity to initial conditions.}
\label{fig:3}
\end{figure}

Next we investigate ECA Rule 30, defined in Fig.~\ref{fig:3}(a), which is categorized as a member of class 3 CA according to Wolfram's universal complexity classes. 
These are CA that produce chaotic and seemingly random patterns, although some small-scale structures are present~\cite{wolfram1984universality}. Remarkably, ECA Rule 30 is one of the simplest known systems to exhibit chaos~\cite{wolfram1986random}. We experimentally demonstrate such a chaotic behaviour of ECA Rule 30 on the same simple photonic hardware in Fig.~\ref{fig:3}(b) starting from a random initial condition. In this case, there is greater variability in the peak pulse intensities compared to Rule 90 due to the lower interference visibility between three optical delay lines. However, the optoelectronic thresholding is still adequate to ensure the intended operation of ECA Rule 30. A necessary (but not sufficient) condition for chaos is sensitivity to initial conditions. Fig.~\ref{fig:3}(c) shows the space-time diagram starting from the same initial condition as Fig.~\ref{fig:3}(b), but with one cell inverted. The region of differences between the two patterns grows linearly to the right with Lyapunov exponent $\lambda_{R}=1$ and asymptotically linearly to the left with Lyapunov exponent $\lambda_{L}\approx0.24$, hence implying an exponential divergence in the cell configurations over time and sensitivity to initial conditions. Other necessary conditions for chaos such as non-periodicity and topological mixing have also been verified empirically~\cite{wolfram1986random}. Due to the simplicity of ECA Rule 30, it can be used as an efficient pseudo-random number generator. This can be accomplished, for example, by taking the sequence defined by the states of the central cell as it evolves in time, i.e. the middle column of the space-time diagram. Therefore, the initial condition acts as the seed. Importantly, ECA Rule 30 is highly nonlinear and computationally irreducible, unlike ECA Rule 90, which is linear (modulo 2) and amenable to algebraic analysis~\cite{martin1984algebraic}. Indeed, detailed statistical analysis of the sequences produced by ECA Rule 30 shows that it is both a fast and safe random number generator~\cite{wolfram1986random}. Unlike previous photonic random number generators~\cite{marandi2012all,reidler2009ultrahigh,stefanov2000optical} relying on quantum processes or other continuous fluctuations, our system is pseudo-random, which means it is deterministic and repeatable given the initial seed. This is often useful in practice to reliably reproduce results in applications such as Monte Carlo simulations~\cite{niederreiter1978quasi}, stream ciphers~\cite{rueppel2012analysis}, and generative adversarial networks~\cite{goodfellow2014generative}. We note that ECA Rule 30 was previously demonstrated using free-space optics~\cite{madjarova1997optical}, however, this implementation encoded cells on pixels of 2D liquid-crystal screens, which introduced some redundancy. In contrast, our approach more faithfully implements the 1D lattice for ECA, can be easily extended to an arbitrary number of cells, and is easily programmable to implement more than just a single rule.

\begin{figure}[t]
\includegraphics{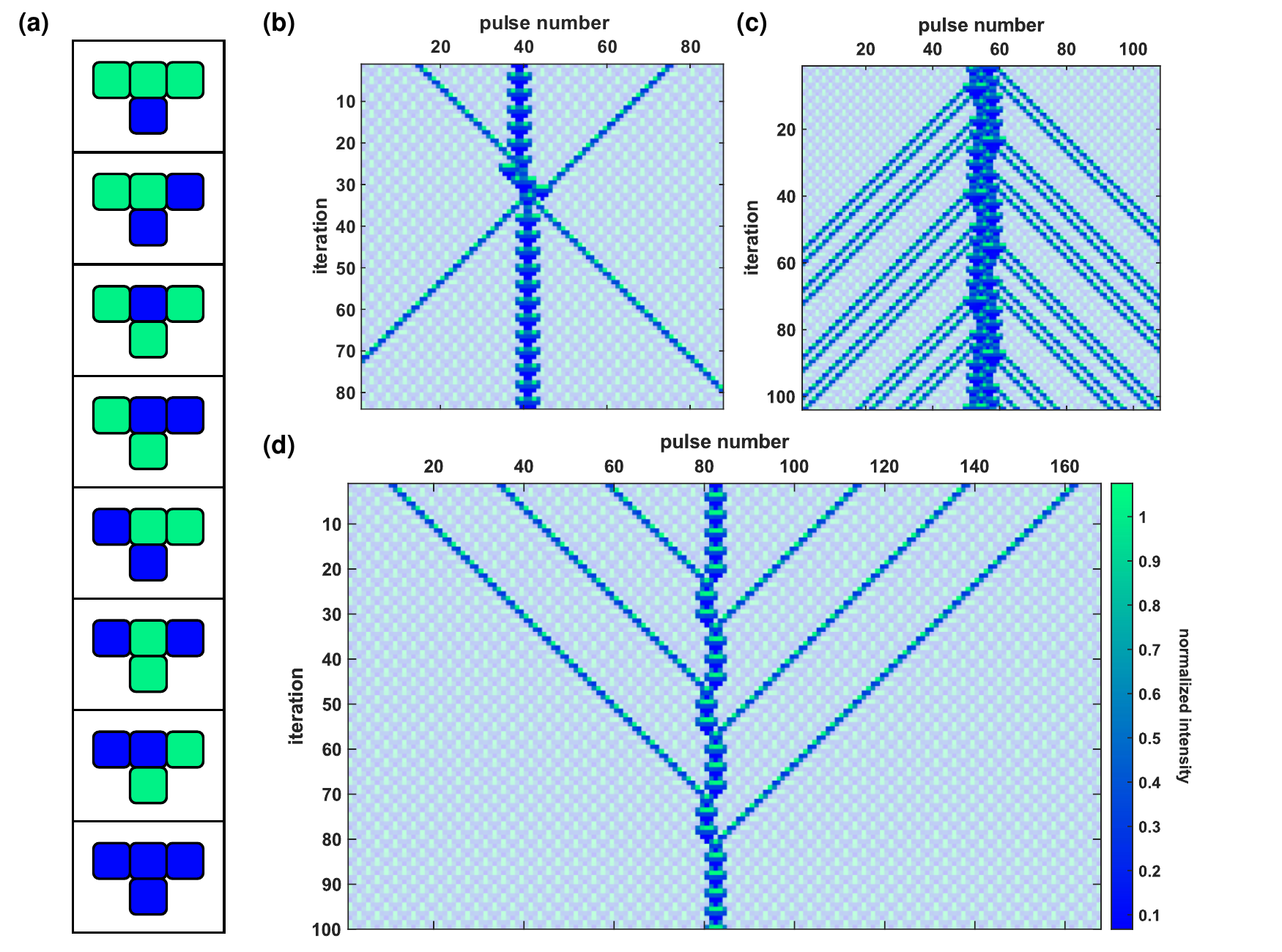}
\caption{\textbf{Soliton-like and glider interactions produced by ECA Rule 54 in the photonic hardware.} (a) Truth table showing the update for ECA Rule 54. Space-time diagrams of ECA Rule 54 with periodic background filtered out, showing (b) glider collision, (c) glider gun, and (d) black hole.}
\label{fig:4}
\end{figure}

Finally, we study class 4 CA, which involve a mixture of order and randomness, with localized structures that move and interact in complicated ways~\cite{wolfram2002new}. A well-studied example of this is ECA Rule 54, defined in Fig.~\ref{fig:4}(a), which can be interpreted as a discrete analogue of excitations in an active nonlinear medium with mutual inhibition~\cite{de2014complete}. In this case, the mobile self-localizations called \textit{gliders} appear on a stable periodic background called the \textit{ether}. Gliders behave like solitons in many regards~\cite{martinez2012soliton}. The study of optical solitons has become a diverse and rich field of study~\cite{agrawal2000nonlinear}. However, optical solitons usually arise due to a balance between nonlinear and linear dispersive effects. In contrast, we have demonstrated optical soliton-like behaviour in a synthetic temporal lattice with only simple binary rules. Despite its simplicity, our system captures physically relevant features since a reversible extension of ECA Rule 54 has produced insightful results in non-equilibrium statistical mechanics and generalized hydrodynamics~\cite{buvca2021rule}. By properly programming ECA Rule 54 in our photonic simulator, we experimentally demonstrated a glider collision, shown in Fig.~\ref{fig:4}(b), whereby gliders emerge after the collision with the same shape and velocity but with a phase shift, which is characteristic of soliton collisions~\cite{zabusky1965interaction}. Such glider collisions can be used to construct logic gates~\cite{martinez2006phenomenology} and Universal Turing Machines~\cite{cook2004universality} for unconventional computing. Furthermore, we also observed a glider gun, shown in Fig.~\ref{fig:4}(c), in which a higher-order localization produces lower-order gliders akin to the process of soliton fission~\cite{wang2004soliton}. Conversely, a glider black hole, shown in Fig.~\ref{fig:4}(d), looks like the process of soliton fusion. Therefore, we have demonstrated a diverse range of glider and soliton interactions in our simple photonic computational platform, which can help unlock new methods of optical information processing.  

\begin{figure}[t]
\includegraphics{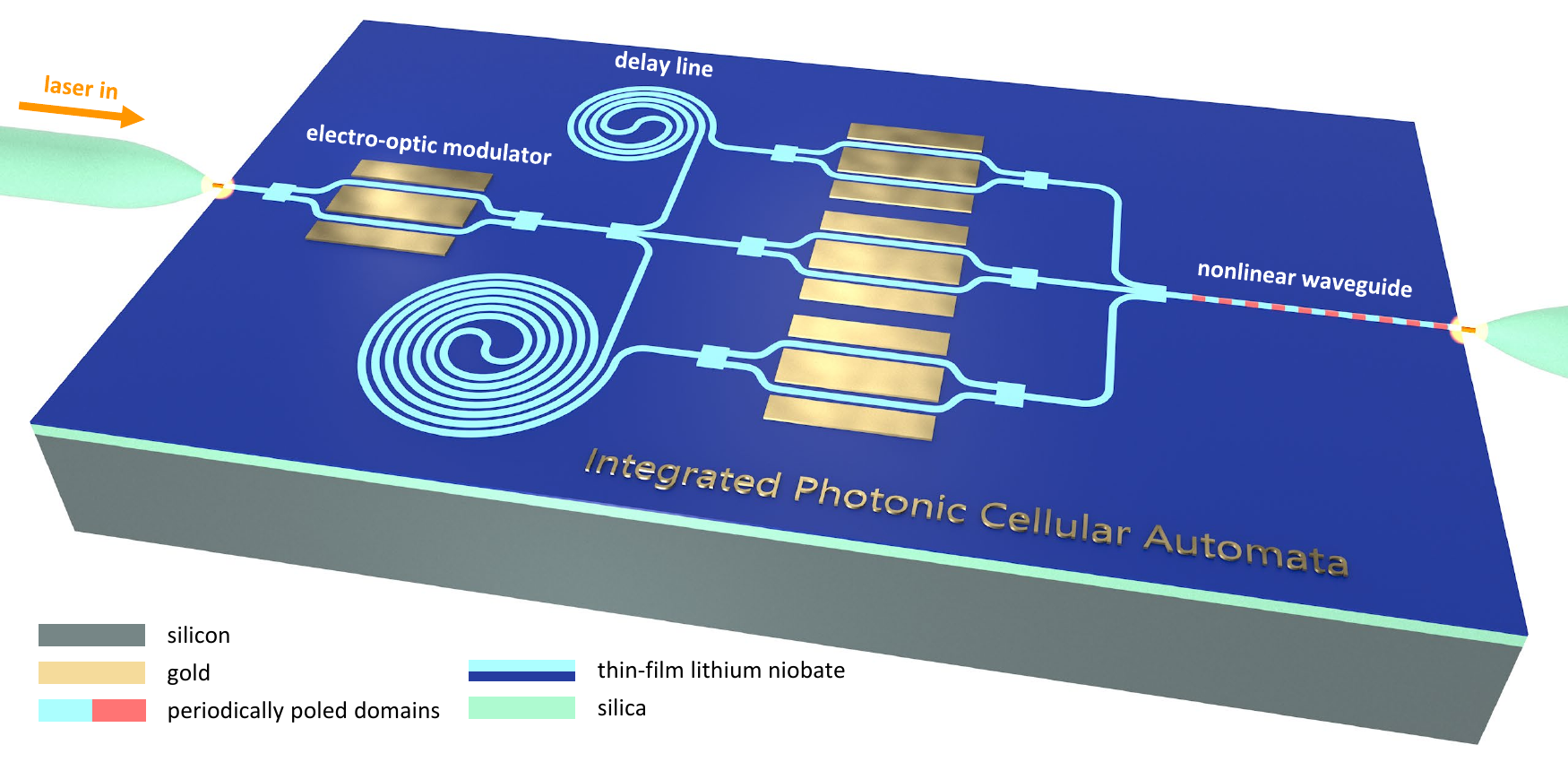}
\caption{\textbf{Lithium niobate nanophotonic cellular automata.} The simplicity of the photonic hardware components for simulating complexity can be maximized by on-chip integration with lithium niobate nanophotonic circuits. For example, integrated EOMs offer greater performance and a periodically-poled nonlinear waveguide can enable efficient all-optical thresholding and feedback. This simple nanophotonic circuit can yield orders of magnitude improvement in the speed and energy-efficiency for simulating complexity in CA.}
\label{fig:5}
\end{figure}
An example on-chip implementation of photonic CA is shown in Fig.~\ref{fig:5} based on the same time-multiplexed architecture of our current experiment, but on a monolithic thin-film lithium niobate platform~\cite{zhu2021integrated} which can increase speed and energy efficiency by potentially orders of magnitude. For instance, in our current implementation we perform the nonlinear activation optoelectronically. However, the photodetector bandwidth ultimately bottlenecks the clock speed of CA. All-optical methods for nonlinear activations can be implemented on the same thin-film lithium niobate chip~\cite{li2022all,guo2022femtojoule} followed by an optical feedback loop to enable terahertz clock rates that are unattainable by digital electronics. The VOAs of our current implementation can be replaced by static integrated EOMs to provide greater control and reproducibility in setting amplitude weights for specific rules. Similarly, other photonic components in our system can be replaced by their simpler and higher performance integrated counterparts.

\section*{Conclusion}
In summary, we have demonstrated a special-purpose photonic computational platform utilizing a synthetic temporal dimension and simple hardware components capable of simulating a wide range of complex phenomena. Simple rules based on local shift-invariant interactions are used to effectively implement different ECA. Our decentralized and non-von Neumann photonic computer can be programmed to represent different rules and initial conditions for the light pulses due to the flexibility and rapid reconfigurability afforded by our hardware system. A range of important complex phenomena including fractals, chaos, and solitons are shown on the same hardware. Future work can involve generalizing the time-multiplexed photonic network to implement other types of CA including filter CA~\cite{park1986soliton}, reversible CA~\cite{buvca2021rule}, and stochastic CA~\cite{goltsev2010stochastic}. This can enable the study of experimentally-challenging complex dynamics in kinetic critical phenomena~\cite{grassberger1984new}, Ising models~\cite{domany1984equivalence}, and lattice Boltzmann models~\cite{wolf2004lattice}. Furthermore, achieving complexity from simple photonic hardware is an important first step towards harnessing this complexity for efficient and robust artificial intelligence, for example in reservoir computing~\cite{yilmaz2014reservoir,nichele2017deep} and deep learning~\cite{mordvintsev2020growing,randazzo2020self}. Our results can inspire a path for special-purpose photonic computers enabling ultrafast low-power operation for critical real-time and edge-computing applications, and new information processing strategies using light.

\section*{Methods}
\subsection*{Experimental setup}
For a more detailed picture of our experimental setup, please see Supplementary Information Fig.~1. A mode-locked laser (MLL) is used that outputs femtosecond optical pulses with a center wavelength of $1550~{\rm nm}$ and a repetition period of $T_{\rm R}=4~{\rm ns}$. Then, the pulses are stretched to $\sim\!5~{\rm ps}$ with a $200~{\rm GHz}$ Channel 34 filter to reduce the effects of dispersion. After the pulses are stretched, $10\%$ of the power is tapped with a 90:10 optical fiber splitter and sent directly to a $600~{\rm MHz}$-bandwidth photodetector. The RF output of the detector passes through a $300~{\rm MHz}$ low pass filter, which isolates the $250~{\rm MHz}$ component of the signal. This signal acts as a clock for the FPGA (Zynq UltraScale+ RFSoC), which generates the modulator driving signals for the EOM in the experiment. Deriving the FPGA's clock directly from the optical pulse train eliminates any timing drift between the optical path and electronic signals. The $90\%$ of the optical power that is not used to clock the FPGA is instead sent through two consecutive intensity modulators (IMs). The first IM, converts the uniform input pulse train to a binary string that contains either an initial condition or the previous state of the ECA under study. The second IM, helps to achieve a better extinction ratio for the zeros in these binary strings. After exiting the modulators, the binary pulse train passes through an erbium-doped fiber amplifer (EDFA) and another $200~{\rm GHz}$ Channel 34 filter. Then, pulses are first split between two paths at a 50:50 spliter. One of these paths leads to a second 50:50 splitter, where the pulses are again divided between another two paths. The paths after the second 50:50 splitter are labeled the $\pm1T_{\rm R}$ delay lines. The lengths of these lines are chosen to delay advance the pulse train by one repetition period relative to the $0T_{\rm R}$ delay line, which is the other line after the first 50:50 splitter. The result of delaying and advancing the pulse train in this manner is coherent interference nearest-neighbour pulses once the delay lines are recombined. To detect the state, the output pulse train passes through another EDFA and $200~{\rm GHz}$ Channel 34 filter. The pulses are split at a final 50:50 splitter and the signal is then measured on both a fast 5 GHz-bandwidth photodetector and a slow kHz photodetector. The RF output of the slow detector is sent to the stabilization electronics for the delay lines, whilst the RF output of the fast detector is recorded on an oscilloscope. The optoelectronic signal is thresholded electronically to produce binary states, which are then sent to and stored on the FPGA, which uses a digital-to-analog converter (DAC) to convert the array into an RF pulse pattern for the next ECA iteration.
\subsection*{Programming elementary cellular automata rules}
Setting the desired ECA rule involves adjusting both the relative intensities and phases between the three delay lines. VOAs are used to adjust the intensities in the lines by detuning the coupling in the free space delays shown in Supplementary Information Fig.~1, and the the relative phases are set to either $0$ or $\pi$ by changing the feedback signals from the PIDs used to stabilize the $\pm1T_{\rm R}$ delay lines. A relative phase of $0$ represents constructive interference between two delay lines, and conversely a relative phase of $\pi$ represents destructive interference. Therefore, the result of the $\pm 1T_{R}$ delay lines, tuning the VOAs, and setting relative phases can be summarized as:
\begin{equation*}
    y_{i}(t) = a_{-1}x_{i-1}(t)+a_{0}x_{i}(t)+a_{1}x_{i+1}(t)\ ,
\end{equation*}
where $x_{i}(t)$ is the amplitude of the $i^{th}$ light pulse in the $t^{th}$ iteration before being split into the delay lines, $y_{i}(t)$ is the amplitude of the light pulse after recombining delay lines, and $\{a_{-1},a_{0},a_{1}\}\in[-1,1]$ are the losses set by the VOAs and phases representing constant linear weights. The light pulse amplitude $y_{i}(t)$ is converted to an intensity $|y_{i}(t)|^{2}$ after passing through the photodetector and then optoelectronic thresholding performs the function:
\begin{equation*}
    x_{i}(t+1) = H\left(|y_{i}(t)|^{2}-b\right)\ ,
\end{equation*}
where $H(x)$ is the Heaviside step function, $b$ is the thresholding value, and $x_{i}(t+1)$ is the output result to be reinjected as the light pulse amplitude for the next iteration. Therefore, any light intensity $|y_{i}(t)|^{2}<b$ represents a dead cell, and conversely any light intensity $|y_{i}(t)|^{2}>b$ represents a live cell. The particular mappings for each ECA rule studied in the Results section is given in Supplementary Information Section 3.
\subsection*{Acknowledgments}
The authors gratefully acknowledge support from ARO grant no. W911NF-18-1-0285, NSF grant no. 1846273 and 1918549, AFOSR award FA9550- 20-1-0040, and NASA/JPL. The authors wish to thank NTT Research for their financial and technical support.

\subsection*{Author Contributions}
All authors contributed to this manuscript.
\subsection*{Competing Interests}
The authors declare no competing interests.
\subsection*{Data Availability}
The data used to generate the plots and results in this paper are available from the corresponding author upon reasonable request.
\subsection*{Code Availablility}
The code used to analyze the data and generate the plots for this paper is available from
the corresponding author upon reasonable request.
\bibliography{apssamp}
\newpage

\end{document}